# Size-dependent enhancement of passive microwave rectification in magnetic tunnel junctions with perpendicular magnetic anisotropy

A. Sidi El Valli[1a], V. Iurchuk[1,b], G. Lezier[1], I. Bendjeddou[2], R. Lebrun[3], N. Lamard[1], A. Litvinenko[1], J. Langer[4], J. Wrona[4], L. Vila[1], R. Sousa[1], I. L. Prejbeanu[1], B. Dieny[1], and U. Ebels[1a]

1. Univ. Grenoble Alpes, CEA, CNRS, Grenoble INP*, IRIG-Spintec, 38000 Grenoble, France

2. *RFIC-Lab - Univ. Grenoble Alpes, Grenoble, France*

3. *Unité Mixte de Physique, CNRS, Thales, Université Paris-Saclay, Palaiseau, France*

4. Singulus Technologies AG, 63796 Kahl am Main, Germany

[a] Corresponding authors: ahmed.sidielvalli@cea.fr, ursula.ebels@cea.fr

[b] Current address: Institute of Ion Beam Physics and Materials Research, Helmholtz-Zentrum Dresden-Rossendorf, 01328 Dresden, Germany

## Abstract

Spintronic *rf* detectors were demonstrated, recently, for energy harvesting and wireless communication at low input power. Here we report on the optimization of the rectified output *dc* voltage using magnetic tunnel junctions (MTJ) with strong perpendicular anisotropy (PMA) of both the polarizing and the free layer. The magnetization of the polarizing layer is fixed out-of-plane, while the free layer thickness is adjusted so that its magnetization orientation changes from in-plane to out-of-plane. The rectification *dc* output voltage lies in the mV range for moderate *rf* powers, with a signal-to-noise ratio of 10-100 for $P_{rf}$ = –25dBm. It shows a strong dependence on the dimensions of the MTJ: it increases by a factor of 5-6 when reducing the diameter from 150nm to 20nm. This enhancement can be doubled when reducing the FL thickness from 1.8nm to 1.6nm. This dimensional enhancement is attributed to the change of the effective anisotropy of the excited free layer, and the MTJ resistance. The results are of interest for the design of spintronic based *rf* detectors with optimized sensitivity.

Nanoscale magnetic tunnel junctions (MTJ), in addition to their applications in memories and sensors, have a strong potential in wireless communication applications[2,3] and neuromorphic computing[4]. Injecting an *rf* signal in an MTJ, with a frequency close to the ferromagnetic resonance (FMR) frequency of the free (or polarizing) layer, generates a *dc* voltage in the output[5]. This frequency selectivity in the *rf*-to-*dc* conversion makes the MTJ-based nanoscale microwave detector of interest for low power applications [6, 7, 8, 9]. High detection sensitivity exceeding the sensitivity of Schottky diodes, has been reported[10, 11] for MTJs with a uniformly magnetized free layer in zero external magnetic field[9], as well as for vortex based MTJs in the presence of an external magnetic field[11]. Further enhancements of the sensitivity were also possible through parametric synchronization[12]. However, the larger sensitivity required an additional *dc* bias current and, in some cases a high magnetic field, which limits their power efficiency as passive (*dc* bias free) *rf* detectors and energy harvesters.

In this letter, we investigated the passive *rf*-to-*dc* conversion in magnetic tunnel junctions derived from perpendicular MTJ stacks used for magnetic random access memory (MRAM) applications[13, 14]. This study was driven by the question under what conditions memory and *rf* detection functionalities can be obtained for the same MTJ stacks and how to increase the efficiency via the MTJ geometry (free layer thickness and device diameter). Output signal levels in the mV range were obtained for moderate *rf* powers. Reducing the diameter of the MTJ enhanced the output *dc* voltage by a factor of ~ 5 for the same FL thickness. It is further improved when the FL thickness is at the transition from the in- to out-of-plane orientation. The corresponding signal-to-noise ratio (SNR) was larger than 10 at the lowest power levels, reaching values of 90 when decreasing the MTJ diameter from 150nm to 20nm.

The MTJ devices are bottom pinned perpendicular magnetized MTJ stacks with the following composition[15] (thickness in nm), see Fig. 1(a): bottom contact/ [Co 0.5 / Pt 0.2]$_6$ / Ru 0.8 / [Co 0.6 / Pt 0.2]$_3$ / Ta 0.2 / Co 0.9 / W 0.25 / FeCoB 1 / MgO 0.8 / FeCoB (FL) / W 0.3 / FeCoB 0.5 / MgO 0.75nm /top contact. The FeCoB free layer (FL) has three different thicknesses, $t_{FL}$=1.8, 1.6 and 1.4nm. Decreasing the thickness reorients its magnetization from in-plane (1.8nm) to out-of-plane (1.4nm) because of the competition between the interfacial perpendicular magnetic anisotropy (iPMA) at the FeCoB/MgO interface and the demagnetizing energy. The magnetization of the composite Co/W/FeCoB polarizing layer PL is oriented out-of-plane due to the iPMA at the MgO interface and the interlayer exchange coupling to the (Co/Pt)$_6$/Ru/(Co/Pt)$_3$ synthetic antiferromagnet (SAF) structure. The magnetic stacks were patterned into circular nanopillars using e-beam lithography followed by Ar+ ion etching. The nominal diameters D are D = 20, 40, 80, 100 and 150nm. Statistical measurements on the fabricated devices at low bias voltage give tunneling magneto-resistance ratios (TMR) of 100%. The resistance area product of the MTJ stack is close to 10Ω·μm². It is noted here, that the MTJ stacks were deposited on CMOS wafers as part of a monolithic CMOS-spintronics co-integration test[15]. In this case the bottom electrode material was TaN, which leads to a relatively high resistance of $R_s \sim 200\ \Omega$ in series with the MTJ resistance.

The orientation of the free layer was verified via PPMS measurements of devices of 100nm diameter. Corresponding magnetoresistance (MR) loops for magnetic fields applied out-of-plane $H_\perp$ are shown in Fig. 1(b) and for in-plane fields $H_\parallel$ in Fig. 1(c). Under $H_\perp$ (Fig. 1(b)) the transition from the square hysteresis loop (t=1.4nm) to an S-shaped loop (t=1.8nm) confirms the reorientation of the free layer magnetization from out-of-plane to in-plane respectively. It also shows an offset field of $\sim 45 mT$ arising from a non-compensated SAF. Under $H_\parallel$ (Fig.1 (c)) the MR changes are weak, for $t_{FL} = 1.8$nm and $t_{FL} = 1.6$nm the decrease of MR with field arises from the gradual rotation of the PL towards $H_\parallel$. For $t_{FL} = 1.4$nm, under $H_\parallel$, both the FL and PL are out-of-plane and their relative angle changes are weak and lead to a quasi-flat MR.

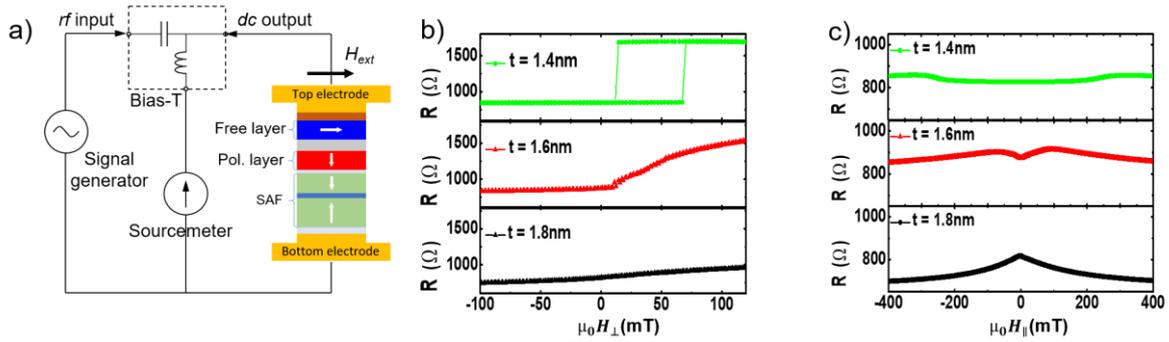

Figure 1 : (a) Schematics of the experiment and the MTJ stack structure. The *rf* and *dc* components are separated using a bias-T. (b) Magnetoresistance of 100nm diameter nanopillars measured for out-of-plane field $H_\perp$ for all FL thicknesses. (c) The same as (b) but with an in-plane magnetic field $H_\parallel$.

The passive detection properties have been investigated under in-plane field for all three thicknesses, focusing on the influence of the device geometry on the output signal. The measurements were conducted at room temperature using a standard spin torque ferromagnetic resonance (ST-FMR) setup (see. Fig. 2(a)) where an *rf* signal is injected through a bias-T to the MTJ device. The *rf* signal induces resonant oscillations of the free (or polarizing) layer magnetization via the spin-transfer torque, and thereby leads to an oscillating magnetoresistance. The time-averaged product of the *rf* current and the oscillating resistance – which results in a *dc* voltage – is measured by a voltmeter. In this experiment, no lock-in amplification was used. Fig. 2(a) and Fig. 2(b) show the rectified voltage when the frequency is swept at constant in-plane field for MTJ devices of D = 80nm and for respectively $t_{FL} = 1.8$nm and $t_{FL} = 1.6$nm. For a FL thickness $t_{FL} = 1.4$nm, under in-plane field, no passive detection was seen.

Rectification signals are observed, when the frequency of the external *rf* source is equal to the ferromagnetic resonance frequency, determined by the applied magnetic field. Two branches exist, a low frequency branch (1-5GHz), and higher frequency branch (8-12GHz). From further analysis (not shown here) the two modes are attributed to FL excitations (low frequency mode) and PL excitations (higher frequency mode). Their frequency field dispersions are given in Fig. 2(c) and Fig. 2(d), revealing the typical Kittel-like increase. In the following, the focus will be on the rectification properties of the FL mode that has higher output signals at zero and low external magnetic fields, and is more suitable for applications for instance in the ISM band of 2.4GHz.

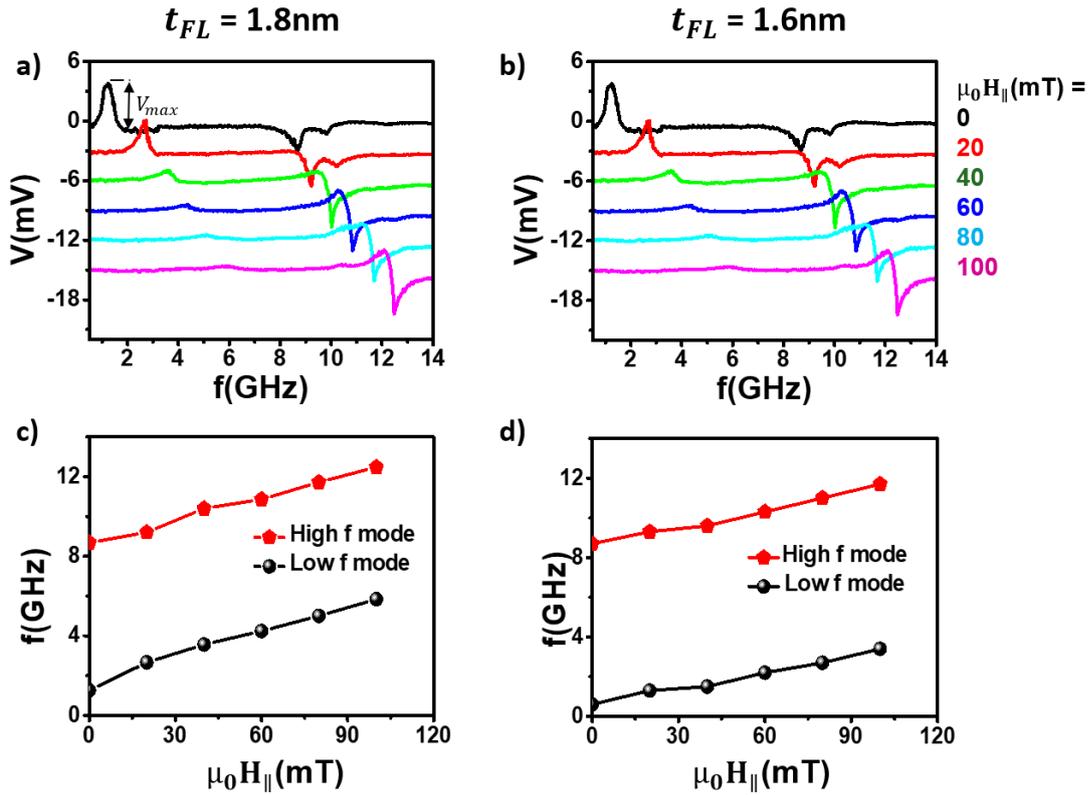

Figure 2 : (a) Passive detection voltage as a function of the input *rf* frequency *f* measured at different in-plane magnetic fields $H_\parallel$ on an MTJ device with D = 80nm diameter and a FL thickness of $t_{FL}$= 1.8nm. The recorded spectra are offset vertically for better visibility. The *rf* source power was -5dBm. On the spectra for $H_\parallel$ =0, the definition of the maximum voltage $V_{max}$, taken at the maximum of the rectification peak, is indicated. (b) The same as (a) but with a FL thickness of $t_{FL}$= 1.6nm. (c) and (d) represent the extracted frequency-field dispersions of the low and high frequency modes, for FL thicknesses of 1.8nm and 1.6nm respectively.

Fig. 3(a) shows the frequency and diameter dependence of the maximum output voltage $V_{max}$ for the FL mode, for devices with a thickness of $t_{FL}$= 1.8nm for which the free layer magnetization is in-plane. The largest *dc* voltage, for all measured diameters, is observed at the lowest frequencies ($\leq 2GHz$). This is a general behavior of resonantly excited FMR modes, because higher frequencies require higher in-plane fields, which, for the FL mode, decreases the amplitude of the magnetization oscillations, and thereby reduces the output voltage. Before we discuss the diameter dependence, we first comment on the signal-to-noise ratio (SNR) that is defined as the signal amplitude $V_{max}$ divided by the *rms* noise background. Fig. 3(b) shows the detection signal $V_{max}$ (solid dots) and corresponding *rms* noise levels (open dots) as a function of the injected *rf* power $P_{rf}$ for $t_{FL}$=1.8nm and the different device diameters. One can see that, when $P_{rf}$ increases from –25dBm to –5dBm, the signal amplitude increases linearly by a factor of 100 from 10µV to 1mV for devices of 150nm diameter and from 100µV to 10mV for devices of 20nm diameter. Moreover for the devices of 80, 100 and 150nm diameter, the noise level is of the order of 1 µV and is almost independent of $P_{rf}$ (only a slight increase is observed for $P_{rf}$ > –10dBm). The devices with 40 and 20nm diameters show higher *rms* noise levels up to 10µV at $P_{rf}$ = –5dBm. A possible origin of the increased noise level is a current-induced heating of the devices with higher resistance (smaller diameter). With this the minimum SNR at $P_{rf}$ = -25dBm is SNR = 22 for D = 150nm and SNR = 90 for D = 20nm.

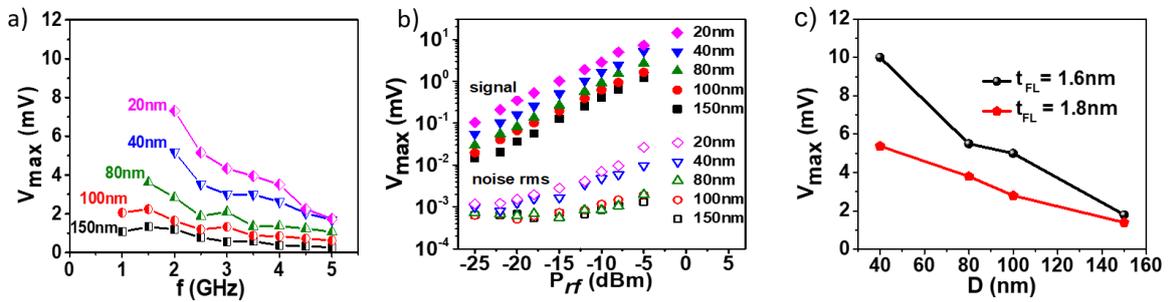

Figure 3: (a) Maximum voltage $V_{max}$ for passive rectification as a function of the *rf* excitation frequency *f*, measured at $P_{rf} = -5$dBm and at zero *dc* bias current for five devices with different diameters (20, 40, 80, 100 and 150nm) and a FL thickness of $t_{FL}$ = 1.8nm. (b) The power dependence of the maximum voltage (closed dots) and the rms of the noise (open dots) of the devices presented in (a) at *f* = 2GHz . (c) Maximum voltage $V_{max}$ for passive detection at *f* = 2GHz as a function of diameter for two different thicknesses $t_{FL}$=1.6 and 1.8nm.

We now turn to the diameter dependence of the output voltage signal when measured at constant *rf* frequency. As seen in Fig. 3(a) and 3(c) for f = 2GHz the output voltage increases upon reducing the diameter D; notably for D=150 nm the maximum output voltage is $V_{max}$ ~1.3mV, and for D = 20nm it is $V_{max}$ = 7.3mV, hence an enhancement by a factor of 5.6. The same diameter dependence was seen for the FL thickness $t_{FL}$ = 1.6nm, but with an overall larger $V_{max}$, see Fig. 3(c) where the maximum voltages as a function of diameter are compared at f = 2GHz. These results indicate that the rectification properties can be optimized by adjusting the free layer thickness and diameter of these MTJs characterized by a strong iPMA of the FL and with an out-of-plane magnetized polarizing layer. In the following, we will investigate the origin of these two dependencies.

The rectification voltage signal is given by the product of the *rf* current $I_{rf}$ and the dynamic resistance change $R(\theta)$, see Eq. 1. The latter is dependent on the amplitude of magnetization oscillations $\delta m_z$, through the projection of the free layer oscillating magnetization onto the polarizer axis, that is considered here to be aligned along the z-axis (out-of-plane). The details of the derivation of Eq. 1 are provided in the supplementary material. The third factor in parenthesis is due to the fact that we consider the cosine angular dependence of the conductance $G(\theta) = R^{-1}(\theta)$[11].

$$V_{dc} = I_{rf} * R(\theta) = I_{rf}(f) * R_p * \left( \frac{1 + TMR}{1 + TMR * \frac{(1 + \delta m_z(j_{rf}, Q, f))}{2}} \right) \quad (1)$$

The oscillating magnetization component $\delta m_z$ can be derived from the linearization of the Landau-Lifshitz-Gilbert (LLG) equation[16], Eq. 2, that includes a damping like ($\sim a_j$) and a field like ($\sim b_j$) torque term, responsible for the excitation of the FL oscillations.

$$\frac{\partial \boldsymbol{m}}{\partial t} = -\gamma(\boldsymbol{m} \times \boldsymbol{H}_{eff}) + \alpha \boldsymbol{m} \times \frac{\partial \boldsymbol{m}}{\partial t} - \gamma a_j j_{rf} \boldsymbol{m} \times (\boldsymbol{m} \times \boldsymbol{P}) - \gamma b_j j_{rf} \boldsymbol{m} \times \boldsymbol{P} \quad (2)$$

Here $\boldsymbol{m}$ is the normalized magnetization of the FL, $\boldsymbol{H}_{eff}$ is the effective field, $\boldsymbol{P}$ is the direction of the magnetization of the PL. The effective field $\boldsymbol{H}_{eff}$ has contributions from the uniaxial interface anisotropy field $\mathbf{H}_a$ due to the iPMA, the demagnetization field $\mathbf{H}_d$ and the Zeeman field $\mathbf{H}_z$. The latter has two contributions, an in-plane field $H_\parallel$ that is varied and a constant out-of-plane field $H_\perp$ due to the stray field of the non-compensated SAF. Through the demagnetization tensor components $N_x$, $N_z$, the demagnetization field depends on the free layer thickness and diameter. In particular, an increase of the diameter reduces $N_z$. Furthermore, $\gamma$, $\alpha$, $a_j$, $b_j$, are respectively: the gyromagnetic factor, the damping constant, the damping-like torque and the field-like torque prefactors, where $a_j$ is defined here as $a_j = \left( \frac{\hbar}{2e} \frac{1}{M_s t} \eta \right)$. The excitation is considered here to occur through the damping-like torque, that after linearization, will provide the susceptibility and through this the oscillating magnetization component $\delta m_z(j_{rf}, Q, f)$. It is given in Eq. 3 (symmetric lorentzian), as a function of the internal magnetic fields, the excitation current density $j_{rf} = \frac{I_{rf}}{Area}$, and the excitation frequency $f$ (see supplementary material for details).

$$\delta m_z(j_{rf}, Q, f) = j_{rf} * [(a_j * (m_{z_0}^2 - 1)] *$$

$$\left[ \frac{\gamma^2 * \alpha * (H_d(Q-1)(2m_{z_0}^2 - 1) + H_\perp m_{z_0} + 2*H_\parallel m_{x_0}) * (f)^2}{([\gamma^2 * (H_d(Q-1)(2m_{z_0}^2 - 1) + H_\perp m_{z_0} + H_\parallel m_{x_0})(H_\parallel m_{x_0})] - (f)^2)^2 + (f)^2 * [\gamma \alpha [H_d(Q-1)(2m_{z_0}^2 - 1) + H_\perp m_{z_0} + 2H_\parallel m_{x_0}]]^2} \right]$$

(3).

$m_{x_0}, m_{z_0}$ represent the static equilibrium position of the magnetization; they are obtained from the energy minimization, calculated analytically, for each external field value.

In Eqs. 2, 3 we have introduced the Q-factor defined as the ratio of the anisotropy field constant and the demagnetization field $Q = \frac{H_a}{H_d}$, that is close to but smaller than one for $t_{FL}$=1.6 and 1.8nm. The anisotropy and demagnetization field depend on the thickness and on the diameter. Hence, varying the diameter or the thickness will lead, via the Q-factor, to a change of $\delta m_z$, and with this to a voltage signal change.

In Fig. 4(a) we have estimated from Eq. 3, the change of the voltage signal due to the diameter change. The values for the parameters $M_s$, $\gamma$, $\alpha$, and iPMA were obtained by fitting the FL FMR mode and are given in the caption of Fig. 4. The values for $N_x$, $N_y$, and $N_z$ were calculated numerically. For details see the supplementary material.

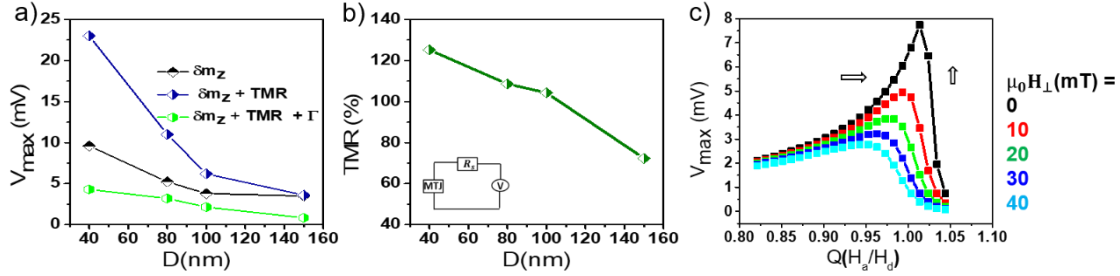

Figure 4: (a) The calculated rectification voltage $V_{max}$ vs. the MTJ diameter obtained from Eq. 1. The contribution due to only the variation of $\delta m_z$, Eq. 3, is shown by the black curve. The blue curve considers as well the diameter dependence of the measured $TMR$, while the green curve takes in addition the impedance mismatch into account via the measured reflection coefficient $\Gamma$. (b) Experimentally measured TMR as function of nominal device diameter. (c) Calculated rectification voltage $V_{max}$ vs. the Q factor for different thicknesses and diameters, and different out-of plane field values. The parameters used in the calculations for (a), (c) are: $M_s$ = 1000kA/m, $\gamma$ = 28GHz/T, $\alpha$ = 0.05, $a_j$ =$7e^{-4}$T/V (D = 150nm), $K_i$ = 873 µJ/$m^2$; $N_x$, $N_y$, $N_z$ are determined numerically for each D and $t_{FL}$. The field values were: $\mu_0 H_\parallel$ = 26mT and $\mu_0 H_\perp$ = 40 mT(for Fig. 4(a)) and $P_{rf}$ = $-5dBm$.

First, when keeping the TMR and *rf* current constant, the tuning of the oscillation amplitude $\delta m_z$ via the diameter leads to an enhancement of the voltage by a factor of ~ 3 when reducing the diameter from 150nm to 40nm (black line in Fig. 4(a)). This increase is explained by two factors. Firstly, at constant RF power, the STT excitation efficiency increases since the current density increases as $\propto \frac{1}{D^2}$. Secondly, the reduction of the demagnetization field $H_d$ increases the Q-factor, and brings the FL magnetization closer to the reorientation transition from in- to out-of-plane where the anisotropy and demagnetization fields compensate. As a result of the two contributions ($j_{rf}$ and Q-factor), the amplitude $\delta m_z$ of the FL magnetization oscillation increases (see Eq. 3). However, this enhancement obtained through the magnetization dynamics and current density accounts only for a portion of the total enhancement observed experimentally. From Eq. 1, it can be seen that the voltage depends also on the TMR which should be independent of the diameter, but due to the serial resistance $R_s$ (~200Ω) from the TaN electrodes the nominal TMR drops for larger diameters, see Fig. 4(b), thereby reducing the voltage. The measured TMR is defined as $\frac{R_{AP}-R_P}{R_P+R_S}$, for small diameters where the MTJ static resistance ($R_p$) is becoming larger than $R_s$ and the effect of the serial resistance is reduced. When including this effective TMR diameter dependence in Eq.1, the total enhancement is ~ 6 (see Fig. 4(a), blue line), indicating that larger MTJ static resistance plays an important role in overcoming the serial resistance effect. However, the diameter dependence of the resistance will also determine the actual *rf* current (or power) that the MTJ device is seeing, due to impedance mismatch. Notably, the increase of the resistance from 550Ω to 2140Ω, when reducing the diameter, see Table 1, increases the *rf* reflections. The *rf* current transmitted to the device $I_{rfdevice}$ can be estimated from the reflection coefficient Γ with $I_{rfdevice} = I_{rf} * \sqrt{1 - \Gamma^2}$, where Γ is obtained from deembedding measurements[17] using a vector network analyzer (see Table 1). When the reflection coefficients are included in the simulations for Eq.1, the total enhancement drops from ~ 6 to ~ 5 (see Fig. 4(a), green line).

| D (nm) | $R_0$(Ω) | Γ @ 2 GHz | $(1 - \Gamma^2)$ |
|---|---|---|---|
| 150 | 550 | 0.7 | 51 % |
| 100 | 870 | 0.8 | 36 % |
| 80 | 1180 | 0.84 | 29 % |
| 40 | 2140 | 0.9 | 19 % |

Table 1. Measured values of the zero-field resistance $R_0$, reflection coefficient Γ, and portion of $P_{rf}$ injected in the devices of different diameters D.

The predicted diameter dependence is in good agreement with the experimental results and shows that the output voltage can be optimized by reducing the device diameter, which reduces the effect of the serial resistance, and for the configuration considered here increases its oscillation amplitude $\delta m_z$ (via the dependence of the susceptibility on the Q-factor and the increase in current density). The reduction due to impedance mismatch could be avoided when introducing an impedance matching network.

In addition to the improvements made through the reduction of diameter the observed increase in signal, when changing the thickness from $t_{FL} = 1.8$nm to $t_{FL} = 1.6$nm is also consistent with the theoretically calculated enhancement by a factor of ~ 1.2 (at *f* = 2GHz, and D = 150nm). At constant *rf* current (or power) and diameter the resistance is the same for both thicknesses (see Fig. 1(b)), hence the voltage enhancement is mostly caused by the magnetization dynamics. A reduced thickness will increase the effective perpendicular anisotropy ~ ($H_a - H_d$) and thus the Q-factor, and will bring the FL magnetization closer to the transition from in- to out-of-plane. This, in consequence increases its oscillation amplitude $\delta m_z$. Moreover, the estimations of the rectified voltage in Fig. 4(c) as a function of the Q-factor show that the magnetization dynamics, and the optimization of the perpendicular stray field $H_\perp$, are the key elements for further optimization of *rf* detectors based on these MTJs with iPMA. By bringing the Q factor closer to one – through the thickness and diameter tuning - the output voltage becomes very sensitive to the Q factor in the range of $0.95 < Q < 1.05$, and to the non-compensated stray field $H_\perp$. For instance, for zero stray field $H_\perp$, and $Q \sim 0.98$, the output voltage can be enhanced by a factor of ~7.5 (~12 for zero in-plane field $H_\parallel$) with respect to the current conditions ($0.82 \leq Q \leq 0.92$).

In conclusion, we have demonstrated passive microwave rectification in MTJ devices with a perpendicular polarizer and a free layer whose orientation can be tuned from in- to out-of-plane with thickness. The output detection dc voltage is in the mV range and was measured without *dc* bias current and at zero or small in-plane magnetic fields ($\mu_0 H_\parallel \leq 25\text{mT}$). We found that the detection voltage increases from 1.3mV to 7.3mV as the diameter of the device decreases from 150nm to 20nm, and the latter improvement is doubled when the FL thickness is reduced from 1.8nm to 1.6nm. Different origins of the observed signal enhancement have been discussed including the variation of the oscillation amplitude $\delta m_z$ via the variation of the Q-factor (through thickness and diameter) and the current density, the electrode serial resistance, as well as the impedance mismatch. It was shown that the improvement with the thickness variation is dominated by the variations of the oscillation amplitude $m_z$, while the enhancement when reducing the diameter is due to the combination of the magnetization dynamics and the electrodes serial resistance. The results are of interest for the optimization of MTJ *rf* detectors for passive detection applications, such as energy harvesting and demodulation in the 2.4GHz IMS band. The measurements also point out the importance of the serial resistance that may occur upon integration of the spintronic devices with CMOS. It demonstrates the advantages of reducing the MTJ size and emphasizes the importance of the consideration of the effective anisotropy, the current density, the static resistance, and the SAF stray field. For applications in higher frequency IMS bands, we propose the exploitation of the PL mode, it has the same diameter dependence and higher output voltages.

**Supplementary Material**

See supplementary material for the complete derivation of the rectification voltage, and the magnetization dynamics.

**Author's contribution**

A.Sidi El Valli, and V.Iurchuk, have contributed equally to this work.

**Acknowledgments**

The authors receive the funding support from the GREAT project (EU Horizon 2020 research and innovation program under grant agreement No 687973). A. Sidi El Valli acknowledges the financial support from the Nanoscience Foundation and the ERC Grant MAGICAL (N°669204).

**Data Availability**

The data that support the findings of this study are available from the corresponding author upon reasonable request.

**References**

[1] T. Chen, R.K. Dumas, A. Eklund, P.K. Muduli, A. Houshang, A.A. Awad, P. Dürrenfeld, B.G. Malm, A. Rusu, and J. Åkerman, Proc. IEEE 104, 1919 (2016).

[2] H.S. Choi, S.Y. Kang, S.J. Cho, I.-Y. Oh, M. Shin, H. Park, C. Jang, B.-C. Min, S.-I. Kim, S.-Y. Park, and C.S. Park, Sci. Rep. 4, 5486 (2014).

[3] X. Liu, K.H. Lam, K. Zhu, C. Zheng, X. Li, Y. Du, C. Liu, and P.W.T. Pong, (2016).

[4] Torrejon, J., Riou, M., Araujo, F. A., Tsunegi, S., Khalsa, G., Querlioz, D., ... & Grollier, J. (2017). Nature, *547*(7664), 428-431.

[5] A.A. Tulapurkar, Y. Suzuki, A. Fukushima, H. Kubota, H. Maehara, K. Tsunekawa, D.D. Djayaprawira, N. Watanabe, and S. Yuasa, Nature 438, 339 (2005).

[6] B. Fang, M. Carpentieri, S. Louis, V. Tiberkevich, A. Slavin, I.N. Krivorotov, R. Tomasello, A. Giordano, H. Jiang, J. Cai, Y. Fan, Z. Zhang, B. Zhang, J.A. Katine, K.L. Wang, P.K. Amiri, G. Finocchio, and Z. Zeng, Phys. Rev. Appl. 11, 014022 (2019).

[7] Marković, D., Leroux, N., Mizrahi, A., Trastoy, J., Cros, V., Bortolotti, P., ... & Grollier, J. (2020). Physical Review Applied, *13*(4), 044050.

[8] Leroux, N., Marković, D., Martin, E., Petrisor, T., Querlioz, D., Mizrahi, A., & Grollier, J. (2021). Physical Review Applied, *15*(3), 034067.

[9] B. Fang, M. Carpentieri, X. Hao, H. Jiang, J.A. Katine, I.N. Krivorotov, B. Ocker, J. Langer, K.L. Wang, B. Zhang, B. Azzerboni, P.K. Amiri, G. Finocchio, and Z. Zeng, Nat. Commun. 7, ncomms11259 (2016).

[10] S. Miwa, S. Ishibashi, H. Tomita, T. Nozaki, E. Tamura, K. Ando, N. Mizuochi, T. Saruya, H. Kubota, K. Yakushiji, T. Taniguchi, H. Imamura, A. Fukushima, S. Yuasa, and Y. Suzuki, Nat. Mater.

[11] Finocchio, G., Tomasello, R., Fang, B., Giordano, A., Puliafito, V., Carpentieri, M., & Zeng, Z. (2021). Perspectives on spintronic diodes. Applied Physics Letters, *118*(16), 160502.

[12] Jenkins, A. S., Alvarez, L. S. E., Freitas, P. P., & Ferreira, R. (2020). Scientific reports, *10*(1), 1-7. 13, 50 (2014).

[13] D. Tiwari, N. Sisodia, R. Sharma, P. Dürrenfeld, J. Åkerman, and P.K. Muduli, Appl. Phys. Lett. 108, 082402 (2016).

[14] W. Skowroński, S. Łazarski, P. Rzeszut, S. Ziętek, J. Chęciński, and J. Wrona, J. Appl. Phys. 124, 063903 (2018).

[15] Chavent, A., Iurchuk, V., Tillie, L., Bel, Y., Lamard, N., Vila, L., ... & Prenat, G. (2020). Journal of Magnetism and Magnetic Mater

[16] Ebels, U., Houssameddine, D., Firastrau, I., Gusakova, D., Thirion, C., Dieny, B., & Buda-Prejbeanu, L. D. (2008). Physical Review B, *78*(2), 024436. 166647.

[17] Koolen, M. C. A. M., Geelen, J. A. M., & Versleijen, M. P. J. G. (1991, September).In Proc. Bipolar Circuits Technol. Meeting (pp. 188-191).

[18] Slonczewski, J. C. (1989). Conductance and exchange coupling of two ferromagnets separated by a tunneling barrier. *Physical Review B*, *39*(10), 6995.